\documentclass[%
 aip,
 amsmath,amssymb,
 reprint,%
]{revtex4-1}

\usepackage{graphicx}
\usepackage{dcolumn}
\usepackage{bm}

\usepackage[utf8]{inputenc}
\usepackage[T1]{fontenc}
\usepackage{mathptmx}

\begin{document}

\preprint{AIP/123-QED}

\title{Enhanced Emission from Ultra-Thin Long Wavelength Infrared Superlattices on Epitaxial Plasmonic Materials}

\author{L. Nordin}
\author{K. Li}%
\author{A. Briggs}
\affiliation{ Department of Electrical and Computer Engineering, The University of Texas at Austin, Austin, TX 78712, USA}%
\author{E. Simmons}
\affiliation{Department of Physics and Applied Physics, UMass Lowell, One University Ave, Lowell, MA 01854, USA}%
\author{S. Bank}
\affiliation{ Department of Electrical and Computer Engineering, The University of Texas at Austin, Austin, TX 78712, USA}%
\author{V.A. Podolskiy}
\affiliation{Department of Physics and Applied Physics, UMass Lowell, One University Ave, Lowell, MA 01854, USA}%
\author{D. Wasserman}
\affiliation{ Department of Electrical and Computer Engineering, The University of Texas at Austin, Austin, TX 78712, USA}%
\email{dw@utexas.edu.}

\date{\today}

\begin{abstract}
Molecular beam epitaxy allows for the monolithic integration of wavelength-flexible epitaxial infrared plasmonic materials with quantum-engineered infrared optoelectronic active regions. We experimentally demonstrate a six-fold enhancement in photoluminescence from ultra-thin (total thickness $\lambda_{o}/32$)  long wavelength infrared (LWIR) superlattices grown on highly doped semiconductor ‘designer metal’ virtual substrates when compared to the same superlattice grown on an undoped virtual substrate. Analytical and numerical models of the emission process via a Dyadic Green’s function formalism are in agreement with experimental results and relate the observed enhancement of emission to a combination of Purcell enhancement due to surface plasmon modes as well as directionality enhancement due to cavity-substrate-emitter interaction. The results presented provide a potential path towards efficient, ultra-subwavelength LWIR emitter devices, as well as a monolithic epitaxial architecture offering the opportunity to investigate the ultimate limits of light-matter interaction in coupled plasmonic/optoelectronic materials.
\end{abstract}

\maketitle
The field of plasmonics centers around the generation and manipulation of hybrid electromagnetic/charge density waves supported at metal/dielectric interfaces\cite{Raether}. Plasmonics’ revival as a field of intense scientific interest, approximately two decades ago\cite{EbbesonEOT}, promised a litany of transformational advances in optics, sensing, and optoelectronics\cite{Zia}. The list of much-heralded applications included, but was not limited to, on-chip sub-diffraction limited waveguiding\cite{Berini,Dereux}, higher efficiency photovoltaics\cite{AtwaterNatMat2010,AtwaterOpEx2010}, sub-diffraction-limited lasers\cite{Hill,Noginov,Gwo}, ultra-efficient emitters\cite{Aydin,Hu}, and enhanced sensitivity sensor systems\cite{Kneipp,Emory,Homola,VanDuyne,Stockman}. However, the promised efficiency gains associated with plasmonic enhancement have largely been offset by the intrinsic losses of plasmonic materials\cite{Khurgin}, especially in the already high  optical quality semiconductor platforms that have benefited from decades of research and development investment from the imaging, sensing, and telecom industries. The mid-IR, however, does not suffer from the affliction of extremely efficient emitters; quite the opposite, in fact. At these long wavelengths, a host of non-radiative recombination mechanisms (Shockley Read Hall, Auger, phonon-assisted, trap-assisted tunneling, etc)\cite{Suchalkin,Wraback,ShanerAPL2007,Luryi,RazehgiAPL2014} conspire to severely limit radiative efficiency, with ever more pronounced effect as the wavelength of emission increases. The inherently low efficiency of mid-IR sources, though, offers very real room for improvement, which can potentially be realized with plasmonic materials engineered specifically for the mid-IR. 

While the noble metals (Au, Ag, etc) are the plasmonic materials of choice at visible and near-IR wavelengths, the large negative real permittivity of the noble metals at longer wavelengths results in optical properties more closely resembling those of perfect electrical conductors (PECs) than plasmonic materials\cite{SlawNP2013}.  The PEC-like nature of traditional plasmonic materials in the mid-IR precludes plasmonic phenomena such as subwavelength confinement of propagating or localized modes, and thus many of the proposed benefits associated with plasmonics, including strongly enhanced light-matter interaction.   At these wavelengths, however, highly doped semiconductors,\cite{SLawOpEx2012,ShanerJAP2011} particularly III-V alloys grown by molecular beam epitaxy (MBE), can demonstrate plasmonic behavior. These epitaxial mid-IR ‘designer metals’ typically employ narrow bandgap materials such as InAs or InAsSb, whose small effective masses and potential for high doping concentrations allow for engineered plasma wavelengths across much of the mid-IR\cite{SLawJVSTB,Tournie}.  The experimental reflection spectra associated with representative examples of such plasmonic semiconductor materials are shown in Figure \ref{figure1}(a) (with growth details and parameter extraction listed in the supplemental material, Table S1).  These materials, coincidentally, also serve as the materials of choice for the active regions of a broad range of mid-IR optoelectronic devices such as interband cascade lasers, superlattice-based emitters and detectors, and even nanostructured mid-IR quantum dot materials\cite{IgorJPD,YHZ,Kurtz,Watkins,BoggessJQE2011,Lan}. 

\begin{figure}
\includegraphics[width=7cm]{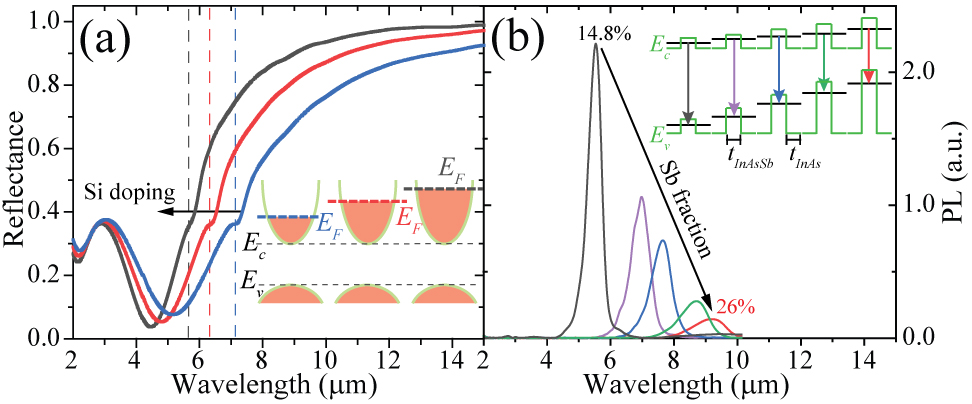}
\caption{\label{figure1} (a) Reflectance spectra from three representative highly doped InAsSb samples, with inset showing a schematic representation of the InAsSb valence band, conduction band, and Fermi level for increasing doping concentration. (b) Photoluminescence spectra for five representative InAs/InAsSb superlattice samples, with inset showing schematic representation of a single period of the superlattices for increasing Sb composition of the InAsSb layer.}
\end{figure}

Of these, MBE-grown semiconductor superlattices (SLs) offer significant design flexibility for engineering absorbers or emitters across the mid-IR. These SLs consist of alternating layers of semiconductor alloys, where the layers are thin enough to allow the overlap of electron (and hole) states in neighboring quantum wells, and thus the formation of minibands in the conduction (and valence) bands, resulting in an effective engineered band-gap for the SL system. When the band offsets of the constituent layers are type-II, either staggered or broken gap, the effective bandgap of the type-II SL (T2SL) can be lower in energy than either of the constituent materials’ bulk bandgaps. Such T2SLs offer the opportunity to engineer absorbing layers with narrow effective band-gaps, and have been the subject of a large amount of interest for both their mid-wave IR (MWIR, 3-5 $\mu$m) and long-wave IR (LWIR, 8-12 $\mu$m) detection capabilities\cite{T2SL,YHZAPL2011,BoggessAPL2012,RazeghiAPL2014T2SL,Chuang,Sanjay}. However, SL materials, before the interest in T2SL detectors, were originally investigated as potential mid-IR emitters, with particular interest in the MWIR wavelength range\cite{YHZ,Kurtz,Baranov}. Unlike band-to-band or type-I emitters, these T2SL-based emitter structures are typically grown relatively thick (1-2 $\mu$m) due to the reduced wavefunction overlap (as electron and hole wavefunctions are largely localized in alternating, adjacent layers). More recently, the significant wavelength flexibility inherent to the SL material system has led to interest in the use of these quantum engineered emitters for the development of LWIR sources\cite{Watkins,BoggessJQE2011}.  Figure \ref{figure1}(b) shows experimental low temperature photoluminescence (PL) from InAs/InAsSb SL emitters as a function of Sb composition, demonstrating this wavelength flexibility. Growth details and emission wavelengths for the T2SLs in Fig. \ref{figure1}(b) are listed in the supplemental material (Table S2).   
The ability to grow both epitaxial plasmonic materials and quantum engineered SL emitters across the LWIR allows for the investigation of coupled emitter/plasmonic systems in a single epitaxial growth.  Control over the spectral position of emission and plasmonic behavior, together with the exquisite uniformity and spatial control offered by MBE, provides a unique system to investigate the near field interaction and enhancement of ultra-thin quantum emitters by plasmonic surfaces.  In this work we demonstrate monolithic integration of an ultra-thin quantum engineered LWIR emitter with an epitaxial plasmonic material, and compare the optical properties of our structure to those of the same emitter, grown on undoped material.

\begin{figure}
\includegraphics[width=7cm]{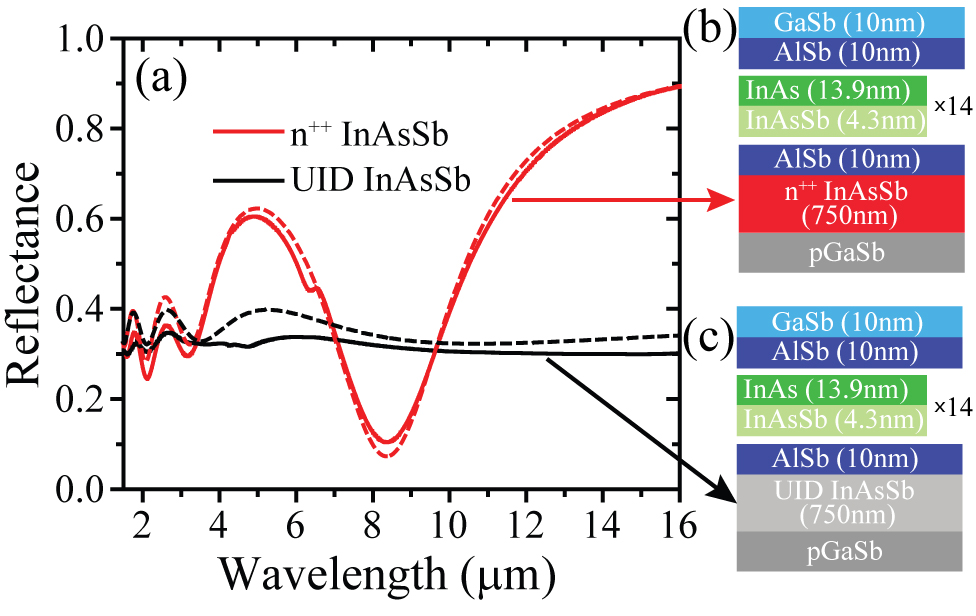}
\caption{\label{figure2} (a) Experimental (solid) and modeled (dashed) reflectance of both the n$^{++}$ (red) and UID (black) InAsSb virtual substrate T2SL emitter samples. The layer structures for the (b) n$^{++}$ InAsSb and (c) UID InAsSb virtual substrates.}
\end{figure}

The sample and control layer structures grown for this work are shown in Fig. \ref{figure2}(b,c).  Our structures are grown by MBE on a p-type GaSb substrate following the growth of a GaSb buffer. The plasmonic (n$^{++}$) structure consists of a $500$ nm thick Si:InAsSb layer, lattice-matched to GaSb, followed by a $255$ nm thick InAs/InAsSb T2SL ($14$ periods), designed for an effective band-gap of $~8.5  \mu$m, bookended by a pair of AlSb carrier blocking layers ($10$ nm each) to promote carrier confinement in the active region.  For the control sample, we grow $500$ nm of unintentionally doped (UID) InAsSb in place of the highly doped (n$^{++}$) InAsSb layer.  The samples are capped with $10$ nm of GaSb to prevent oxidization of the top AlSb layer.  Note that the total thickness of the LWIR emitter active region is only $255$ nm, or approximately $\lambda_{o}/32$, where $\lambda_{o}$ is the free-space wavelength of the band-edge emission from the SL emitter.

Figure \ref{figure2}(a) shows the experimental and fitted reflectance for both the n$^{++}$ and UID virtual substrate samples, normalized to the near perfect reflectance of a Au surface. The plasma wavelength of the n$^{++}$ InAsSb is extracted from reflectance spectra of the as-grown sample. The experimental reflectance spectra are fitted using the transfer matrix method (TMM), treating the emitter as a high index dielectric and the doped layer as a Drude plasmonic material with permittivity:
\begin{equation}
 \epsilon_{o}(\omega)=\epsilon_{\infty}\left(1-\frac{\omega_p^2}{\omega^2+i\gamma\omega}\right)
\end{equation}
with fitting parameters $\omega_{p}$, the plasma frequency, and $\gamma$, the free carrier scattering rate. From the fitting process, the highly doped layer’s plasma wavelength ($\lambda_{p}=2\pi c/\omega_{p}$) is estimated to be $\lambda_{p}=6.7\mu$m and the scattering rate $\gamma=10^{13}$  Hz. X-ray diffraction spectra and further growth details for both samples are provided in the supplementary material (Figure S1). 

\begin{figure}
\includegraphics[width=7cm]{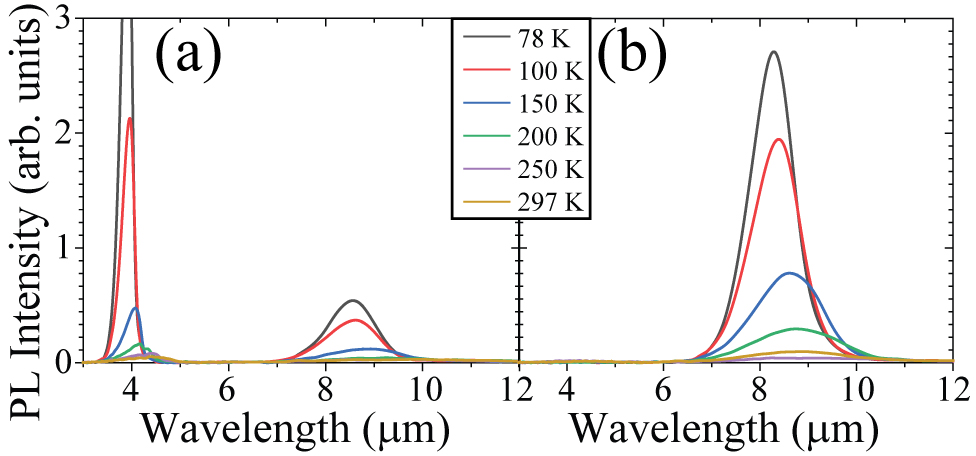}
\caption{\label{figure3} Temperature dependent photoluminescence measured from the ultra-thin T2SL emitters on both the (a) undoped and (b) highly doped virtual substrates.}
\end{figure}

We investigate the emission efficiency enhancement of our ultra-thin plasmonic emitter structure by photoluminescence (PL) Fourier transform infrared (FTIR) spectroscopy, using amplitude modulation step scan mode in order to eliminate the thermal background signal. PL spectra of the n$^{++}$ and UID InAsSb samples are shown in Figure \ref{figure3}, for temperatures from $78$K to $297$K. Comparison of the PL spectra shows a clear (approximately six-fold) enhancement of emission intensity for the T2SL grown above the n$^{++}$ (plasmonic) InAsSb layer, when compared to emission from the T2SL grown on the UID InAsSb virtual substrate. Enhancement is observed for all temperatures and results in observable room temperature emission from the T2SL on the doped substrate, while the room temperature emission from the T2SL on the undoped substrate is at or below the system noise floor. 

\begin{figure}
\includegraphics[width=7cm]{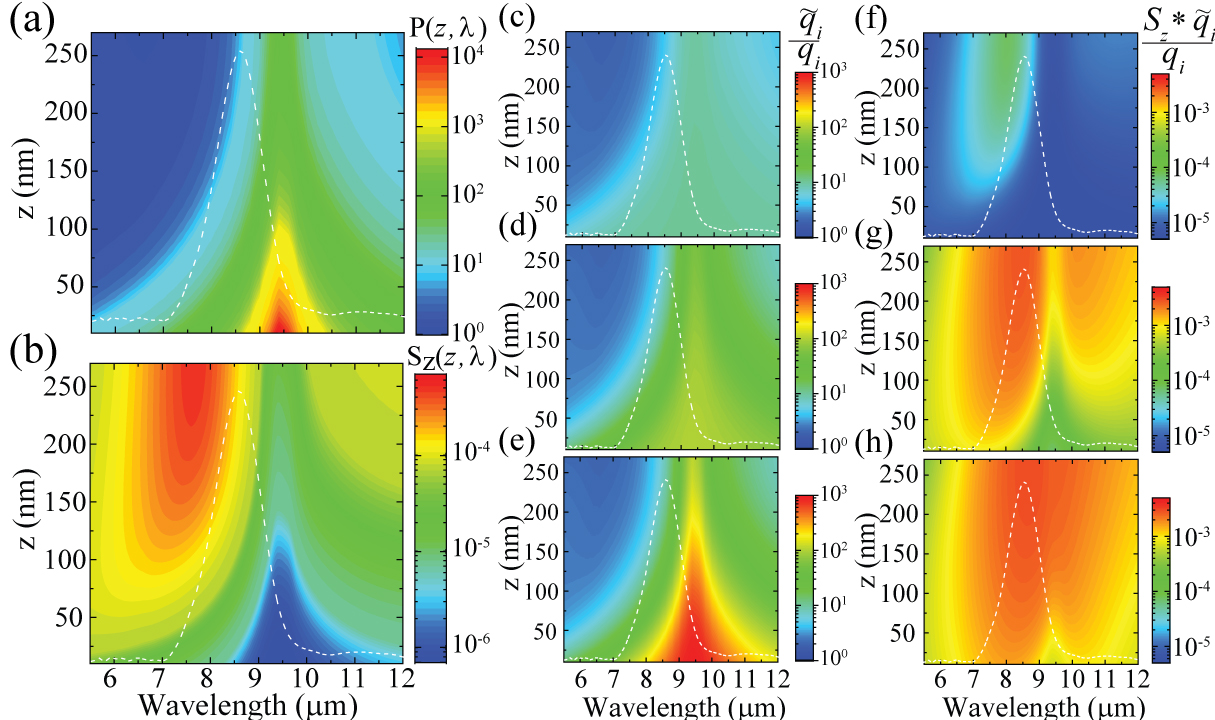}
\caption{\label{figure4} Modelling emission from the LWIR T2SL grown above the n$^{++}$ substrate.  Contour plots of the dipole emitter (a) Purcell factor (P, in logarithmic color scale) and (b) $S_z$ vs. wavelength and position.  Contour plots of the effective quantum yield ($\widetilde{q_i}$, in logarithmic color scale) normalized to the $q_i$ vs. position and wavelength for (c) $q_i=10^{-1}$, (d) $q_i=10^{-2}$, and (e) $q_i=10^{-3}$. The product $S_z*\widetilde{q_i}$ normalized to $q_i$ (in logarithmic color scale) vs. position and wavelength for (f) $q_i=10^{-1}$, (g) $q_i=10^{-2}$, and (h) $q_i=10^{-3}$.  The low temperature PL spectrum from the UID substrate T2SL is overlaid on each plot as a white dashed line.}
\end{figure}

\begin{figure}
\includegraphics[width=7cm]{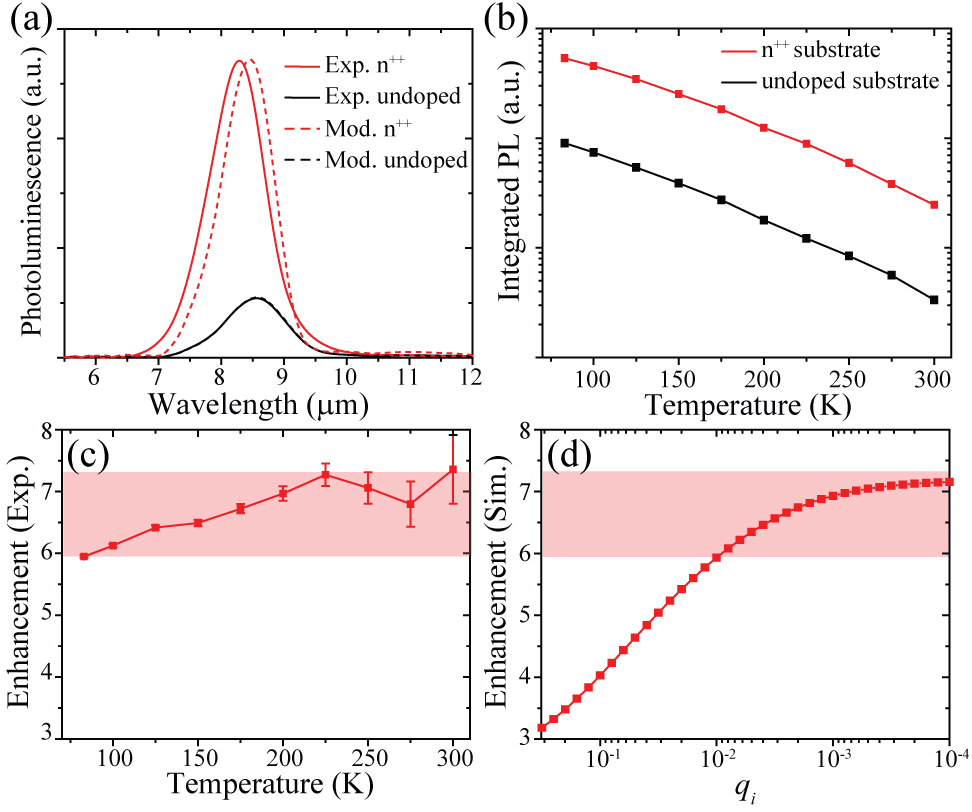}
\caption{\label{figure5} (a) Modeled (dashed, assuming $q_i=0.02$) and experimental (solid) PL from the LWIR T2SL grown on n$^{++}$ (red) and UID (black) virtual substrates.  (b) Integrated PL as a function of temperature for the T2SL on the n$^{++}$ (red) and UID (black) virtual substrates.  An approximately twenty-fold decrease in integrated PL is observed for the samples from T=80K to T=300K. (c) The measured enhancement, $PLE=\int PL_{n^{++}}d\omega / \int PL_{UID}d\omega$, as a function of temperature.  (d) Modeled PLE, for $q_i=0.31$ to $q_i=10^{-4}$.  Shaded regions in (c) and (d) correspond to the range of experimentally observed PLE.}
\end{figure}

The emitters are modelled using a Dyadic Green’s function formalism, incorporated into our TMM. In this approach, the field of a point dipole is first expanded in the (polarization-dependent) plane wave spectrum, parameterized by the components of the wavenumber parallel to the interfaces $k_r$. The TMM formalism is then used to calculate the effect of the reflections within the multi-layer stack\cite{Novotny,Brueck}, followed by the numerical integration of the resulting spectrum. The approach allows for calculation of two inter-related but separate quantities. 
The first of these quantities, given by the imaginary part of the Greens’ function at the origin, represents the (orientation-specific) enhancement of the density of photonic states at the location of the dipole with respect to the density of photonic states in vacuum, a quantity also known as the Purcell factor (P)\cite{Purcell}, 
\begin{equation}
 P=\frac{3}{2}\frac{Im(\vec E \cdot \vec d)}{\omega^2|d|^2}
\end{equation}
with $\vec E$ being the field generated by the point dipole $\vec d$ at the location of the dipole. The enhancement of the density of photonic states modifies the radiative decay rate for the dipole, in turn affecting its intrinsic quantum yield, 
\begin{equation}
 \widetilde{q_{i}}=\frac{Pq_{i}}{1+(P-1)q_{i}}
\end{equation}
where $q_i$ refers to the dipole’s quantum yield (the ratio of the material’s radiative recombination rate to total recombination rate) for the hypothetical isolated dipole in a vacuum. Figure \ref{figure4}(a) illustrates the position- and spectral-dependence of the Purcell factor in the sample with the plasmonic virtual substrate. It is clearly seen that enhancement of the density of photonic states is strongly correlated with the spectral position of the surface plasmon polariton (SPP) mode supported by the planar n$^{++}$ InAsSb/T2SL/air structure ($\lambda_{sp} = 9.4 \mu$m), with the n$^{++}$ InAsSb playing the role of the plasmonic layer. Since the intrinsic quantum efficiency of SL emitters is known to be relatively low\cite{BoggessJQE2011}, the Purcell effect significantly enhances the efficiency of the radiative decay of the T2SL.  Figure \ref{figure4}(c-e) illustrates the effective quantum yield ($\widetilde{q_{i}}$) normalized to ${q_{i}}$ of the dipole emitter within the T2SL for several ${q_{i}}$'s. 
However, the enhancement of the quantum yield alone does not fully describe the overall enhancement of the experimentally observed emission. Because the “added” density of the photonic states come from the guided SPP mode, peaking at $\lambda_{sp}$, the photons emitted into this guided mode are not out-coupled into free space modes and therefore do not contribute to the far-field emission of the overall structure. 
To capture the full experimentally-observed enhancement, we also calculate the $\hat{z}$-component of the Poynting flux density emitted by the dipole ($S_z$), which clarifies the spatial and spectral properties of the emission. In our calculations, the above quantity is normalized to remain independent of the emission frequency and of the local Purcell factor, and includes only emission angles $|\theta|\leq15^o$ from the normal, mimicking our experimental conditions. The behavior of the Poynting flux captures the effect of the directionality/reflectivity reshaping of the emitted light due to the multiple reflections in the optical stack (optical cavity) surrounding the dipole. The distribution of the Poynting flux density as a function of the emission frequency and the location of the dipole within our structures is illustrated in Figure \ref{figure4}(b).  Notably, there is a clear anti-correlation between the Purcell enhancement and the density of modes out-coupled into the far field, which illustrates the fact that the dipoles located close to the metal interface primarily emit into guided (and highly lossy) plasmonic modes. 
The overall enhancement of emission is then proportional to the product of the quantum yield and the cavity-corrected Poynting flux density, $S_z*\widetilde{q_i}$, normalized to the intrinsic quantum efficiency ($q_i$), and shown in Figure \ref{figure4}(f-h). From this product we can see that the spectral regions of maximal enhancement of radiation come from the “compromise” between the maximal Purcell effect and the maximal out-coupling efficiency. Moreover, the enhancement is stronger in systems with smaller $q_i$.
To calculate the final predicted emission spectrum we assume that the point dipole emitters are homogeneously distributed across the T2SL layer, with randomly distributed orientations, and have a distribution of emission frequencies $A(\omega)$ described by the low temperature PL spectrum from the UID substrate T2SL [the black solid line in Fig. \ref{figure3}(a), and the white dashed overlays in Fig. \ref{figure4}]. The overall measurable far-field emission is then given by $S_{tot} = \left< \widetilde{q_i} A(\omega)S_z \right>$,
with $\left< \ldots \right>$ representing an average over the spatial location and orientation of the dipole. The experimental and predicted emission (for  $q_i=0.02$) is shown in Figure \ref{figure5}(a).  As one can clearly see, the calculations predict the observed PL enhancement (PLE) of the emission with remarkable accuracy, and the spectral position of the enhanced emission to within a fraction of a micron.  This minimal discrepancy between modelled and experimental spectra is most likely a result of the slight sample-to-sample variation of our T2SL growths, as the predicted emission from the T2SL on the n$^{++}$ virtual substrate is modeled using the emission from the T2SL on the UID substrate. 

Parameters $\widetilde{q_i}$ and $S_z$ represent two (related) sources of emission enhancement in the multi-layer stacks: efficiency reshaping of emission caused by the Purcell effect and “cavity”-like directionality reshaping caused by the layered environment. While the two parameters are not completely independent, they can be optimized in a semi-independent manner. Therefore, the results presented in this work should be treated as a low-bound estimate of the true potential of the PLE in IR systems that could be improved by, for example, patterning the top surface of the structure. 
Figure \ref{figure5}(b) shows the integrated PL intensity as a function of temperature for the T2SLs on the UID (black) and n$^{++}$ (red) virtual substrates.  As temperature is increased, a dramatic decrease in emission is expected and observed for both samples (an approximately twenty-fold decrease in integrated PL), resulting from increased non-radiative recombination rates (predominantly from increasing Auger recombination)\cite{BoggessJAP2016}.  This decrease in emission intensity can be thought of as a temperature-dependent change in the $q_i$ of the T2SLs, offering a mechanism for exploring the observed and predicted PLE as a function of emitter efficiency.  Thus, the experimental enhancement of emission resulting from the n$^{++}$ virtual substrate is plotted as a function of temperature in Figure \ref{figure5}(c), and we observe a largely monotonic increase in enhancement with increasing temperature (decreasing efficiency).  In Figure \ref{figure5}(d) we plot the calculated PLE as a function of the modeled dipole's $q_i$.  Our model predicts an increase in enhancement with decreasing $q_i$, as expected, with the enhancement saturating at approximately a factor of seven for the structures investigated in this work.  Comparing the experimental PLE to the modeled PLE suggests that our LWIR T2SLs have $q_i$ of ~2$\%$ at low temperature and between 0.02$\%$ and 0.2$\%$ at high temperatures, in line with expected $q_i$ of LWIR T2SL materials\cite{BoggessJQE2011}. Such a decrease in $q_i$ of our emitters agrees with the experimentally observed decrease in integrated PL as a function of temperature observed in Fig. \ref{figure5}(b).  Though correlating the x-axes of Figures \ref{figure5} (c) and (d) would require measurement of the minority carrier lifetimes (not possible due to the weak emission above 100K), the experimental data provides a qualitative agreement with the predictions from theory, and offers insight into the achievable enhancement for LWIR emitters of varying intrinsic efficiencies.  

The mid-IR, and in particular the LWIR, where efficient emitters are severely lacking, offers the opportunity for significant enhancement of emission by combining quantum engineered mid-IR emitters with designer semiconductor plasmonic metals. In this work we demonstrate monolithic integration of an ultra-thin LWIR emitter and epitaxial plasmonic material, as well as the commensurate, six-fold, emission enhancement of the SL emitter relative to the same SL grown on a non-plasmonic material.  We use a theoretical model of the emission based on a Dyadic Green’s function and TMM formalisms to provide a quantitative explanation of the emission enhancement phenomenon as an interplay between Purcell enhancement and emitter-cavity interaction.  Assuming $q_i=0.02$ for our emitters, our model accurately reproduces the emission enhancement observed in experiments. Moreover, the model accurately predicts the increasing PLE for decreasing $q_i$. The LWIR emitters demonstrated in this work open the door to an entirely novel approach to infrared emitter design, where plasmonic materials are implemented, and grown monolithically, with mid-IR active regions for a new class of ultra-thin mid-IR sources. In addition, the ability to engineer both the plasmonic materials’ properties, the active regions’ optical transitions, and (at atomic scale) the geometry of the coupled system, offers a powerful tool-box for the investigation of light-matter interaction with extreme spectral and spatial precision. 
\begin{acknowledgments}
DW, LN, AB and SB gratefully acknowledge support from the National Science Foundation (ECCS-1926187). KL acknowledges support from the National Science Foundation (DMR-1629570). VP and ES acknowledge support from the National Science Foundation (DMR-1629330).
\end{acknowledgments}

\nocite{*}
\bibliography{APLPlasmonicT2SL}
\end{document}